\documentclass[aip,apl,twocolumn,reprint,superscriptaddress]{revtex4-1}
\usepackage{graphicx}
\usepackage{color}
\usepackage{url}

\begin{document}
\title{Temperature-dependent phonon shifts in monolayer MoS$_2$}

\author{Nicholas A. Lanzillo}
\affiliation{Department of Physics, Applied Physics, and Astronomy, Rensselaer Polytechnic Institute, 110 8$^{th}$ Street, Troy, NY 12180}

\author{A. Glen Birdwell}
\affiliation{Sensors and Electron Devices Directorate, US Army Research Laboratory, Adelphi, MD 20723, USA}

\author{Matin Amani}
\affiliation{Sensors and Electron Devices Directorate, US Army Research Laboratory, Adelphi, MD 20723, USA}

\author{Frank J. Crowne}
\affiliation{Sensors and Electron Devices Directorate, US Army Research Laboratory, Adelphi, MD 20723, USA}

\author{Pankaj B. Shah}
\affiliation{Sensors and Electron Devices Directorate, US Army Research Laboratory, Adelphi, MD 20723, USA}

\author{Sina Najmaei}
\affiliation{Department of Mechanical Engineering and Materials Science, Rice University, Houston, TX 77005, USA}

\author{Zheng Liu}
\affiliation{Department of Mechanical Engineering and Materials Science, Rice University, Houston, TX 77005, USA}

\author{Pulickel M. Ajayan}
\affiliation{Department of Mechanical Engineering and Materials Science, Rice University, Houston, TX 77005, USA}

\author{Jun Lou}
\affiliation{Department of Mechanical Engineering and Materials Science, Rice University, Houston, TX 77005, USA}

\author{Madan Dubey}
\affiliation{Sensors and Electron Devices Directorate, US Army Research Laboratory, Adelphi, MD 20723, USA}

\author{Saroj K. Nayak}
\affiliation{Department of Physics, Applied Physics, and Astronomy, Rensselaer Polytechnic Institute, 110 8$^{th}$ Street, Troy, NY 12180}

\author{Terrance P. O'Regan}
\affiliation{Sensors and Electron Devices Directorate, US Army Research Laboratory, Adelphi, MD 20723, USA}

\date{\today}

\begin{abstract}
We present a combined experimental and computational study of two-dimensional molybdenum disulfide (MoS$_2$) and the effect of temperature on the frequency shifts of the Raman-active E$_{2g}$ and A$_{1g}$ modes in the monolayer. While both peaks show an expected red-shift with increasing temperature, the frequency shift is larger for the A$_{1g}$ more than for the E$_{2g}$ mode. This is in contrast to previously reported bulk behavior, in which the E$_{2g}$ mode shows a larger frequency shift with temperature. The temperature dependence of these phonon shifts is attributed to the anharmonic contributions to the ionic interaction potential in the two-dimensional system. 
\end{abstract}

%\keywords{}
\maketitle

%%%%%%%%%%%%%%%%%%%%%%%%%%%%%%%%%%%%%%%%%%%%%%%%%%%%%%%%%%%%%%%%%%%%%%%%
\section{Introduction}
Molybdenum disulfide (MoS$_2$) is transition-metal dichalcogenide consisting of covalently bound S-Mo-S layers held together by the weak van der Waals interaction and is of great research interest due to its potential uses in electronic and optical devices. MoS$_2$ shows a number of interesting features when confined to a single monolayer - most notably an indirect-to-direct electronic band gap transition\cite{mak2010atomically,zuc2011influence,ellis2011indirect} - accompanied by an enormous increase in the photoluminiscence quantum yield\cite{splendiani2010emerging}. The room-temperature moblility of MoS$_2$ is comparable to that of graphene nanoribbons, and has been predicted to be as high as 400 cm$^2$V$^{-1}$s$^{1}$ when optical phonon scattering and intra/intervalley deformation potential couplings are included\cite{kaasbjerg2012phonon}.

The vibrational properties of bulk, few layer and monolayer MoS$_2$ have been studied both experimentally\cite{chakraborty2012symmetry,livneh2010resonant,rice2013raman,kiosegoglou2012valley,lee2010anomalous,bertrand1991surface} and theoretically\cite{molina2011phonons,ataca2011comparative,li2012ideal}. One of the most pronounced effects of confining the system to a strictly 2-dimensional geometry as in the monolayer is the redshift of the Raman-active $A_{1g}$ mode and blueshift of the Raman-active $E_{2g}$ phonon mode with decreasing number of layers\cite{ataca2011comparative}. The unexpected blueshift of the $E_{2g}$ peak was shown to be due to dielectric screening of the long-range Coulomb interaction\cite{molina2011phonons}, while the redshift of the $A_{1g}$ mode is in line with the classical picture of a harmonic potential. Work involving the effect of strain shows that the in-plane $E_{2g}$ mode is sensitive to strain, while the out-of-plane $A_{1g}$ mode shows a weak strain dependence\cite{rice2013raman}. These two Raman-active mode also show distinct doping dependence; with the $A_{1g}$ mode decreasing in frequency with increased electron concentration and the $E_{2g}$ mode showing an overall weak dependence on electron concentration\cite{chakraborty2012symmetry}. This difference is attributed to the stronger coupling to electrons of the $A_{1g}$ mode compared to the $E_{2g}$ mode. 

While the temperature dependence of the Raman active peaks has been investigated for the bulk crystal\cite{livneh2010resonant}, the temperature effect on the monolayer has not been studied. In the bulk crystal, both the $A_{1g}$ and $E_{2g}$ peaks shift toward lower frequencies with increasing temperature, and the temperature coefficient of the $E_{2g}$ mode is larger in magnitude than that of the $A_{1g}$ mode. Here we show using Raman spectroscopy along with finite-temperature molecular dynamics simulations that for the monolayer, both peaks show a redshift with increasing temperature but the magnitude of the frequency shift is larger for the $A_{1g}$ mode than for the $E_{2g})$ mode. To our knowledge, this is the first time that finite-temperature molecular dynamics has been used to study the phonon spectrum of a quantum-confined two-dimensional system. 

%%%%%%%%%%%%%%%%%%%%%%%%%%%%%%%%%%%%%%%%%%%%%%%%%%%%%%%%%%%%%%%%%%%%%%%%
\section{Methods}
\subsection{Theory}
Our molecular dynamics simulations were performed with the CPMD code\cite{cpmd} using HGH gradient-corrected pseudopotentials\cite{goedecker1996separable} and a plane wave cutoff energy of 80 Rydberg. Our supercell contained 48 atoms and k-point sampling was restricted to the $\Gamma$-point of the Brillouin Zone. We have included 14 \AA{} of vacuum separating periodic images of the monolayer. The geometry was relaxed until the forces on the atoms were less than $10^{-2}$ eV/ \AA{}, and we have found the average Mo-S bond length to be 2.41 \AA{}, which is within 0.05 \AA{} of the optimized structures found in previous work\cite{molina2011phonons, ataca2011comparative}. In order to investigate the effect of temperature on the phonon spectra we have performed Car-Parrinello\cite{Car1985unified} MD simulations for the system in the crystal phase. The system was kept close to the Born-Oppenheimer surface through repeated quenching. The simulations were allowed to evolve for 100 ps using a timestep of 4 a.u. and a fictitious electron mass of 400 a.u. The phonon spectrum was derived from the Fourier transform of the velocity auto-correlation function. The resolution in frequency space is on the order of 2 cm$^{-1}$. These calculations were run on 1024 processors of the IBM BG/L supercomputer. 

\subsection{Experiment} 

Single layer MoS$_2$ films were grown directly on 285 nm SiO$_2$/Si substrate by chemical vapor deposition using the procedure described in detail by Najmaei et al.\cite{najmaeiarxiv,zhan2012large}.
Hydrothermally grown MoO$_3$ nanoribbons were dispersed onto auxiliary silicon substrates and placed inside a tube furnace surrounded by the growth substrate. Sulfur powder was sublimated upstream near the opening of the furnace at an approximate temperature of 600$^\circ$C while the furnace was heated to a peak temperature of 850$^\circ$ under a constant flow of nitrogen and was held at this set point for 10 to 15 minutes and then cooled to room temperature. This resulted in incomplete growth of triangular MoS$_2$ domains which were primarily monolayer but occasionally contained a small bilayer region in the center. 

Micro-Raman and photoluminescence (PL) measurements were performed with a WITec Alpha 300RA system using the 532 nm line of a frequency-doubled Nd:YAG laser as the excitation source.  The spectra were measured in the backscattering configuration using a 100x objective and either a 600 or 1800 grooves/mm grating.  The spot size of the laser was ~342 nm resulting in an incident laser power density $~$140 $\mu$W/$\mu$m$^2$. This laser power was found to be sufficiently low to not to cause any shifting in the both the in-plane and out-of-plane modes of the Raman signature\cite{najmaei2012thermal}. Single point Raman measurements were performed on the same location in the sample over the temperature range from 30 to 175$^\circ$C using a heating stage. The location was determined via Raman mapping the selected crystal at each temperature set, after allowing at least 30 minutes for thermal stabilization of the sample and optics. In addition atomic force microscopy (AFM) was utilized to confirm the thicknesses of the CVD material.

%%%%%%%%%%%%%%%%%%%%%%%%%%%%%%%%%%%%%%%%%%%%%%%%%%%%%%%%%%%%%%%%%%%%%%%%
\section{Results and Discussion}

\subsection{Experiment}
Figure 1 shows photoluminescence (PL) and AFM images of a typical monolayer crystal with a small bilayer region at its center.  The AFM images indicate that the deposited MoS$_2$ films are monolayer with a typical thickness of 0.9 nm. In addition, the enhancements of the PL signal emanating from the monolayer regions are indicative of the indirect-to-direct gap transition observed in MoS$_2$ as the layer count is reduced from two or more layers to a single layer\cite{mak2010atomically}. Single spectra recorded at 30 and 175$^\circ$C are shown in \textbf{Figure 2}, and extracted values of the temperature coefficient for the E$^{1}$$_{2g}$ and A$_{1g}$ modes for both CVD grown monolayer and bulk ($<$100 nm thick) exfoliated MoS$_2$ are presented in \textbf{Figure 3.} An increased phonon coupling in the MoS$_2$ is observed at elevated temperatures, as shown in the increased intensity of the peaks for both monolayer (\textbf{Figure 2}) and bulk (not shown).

\begin{figure}[h!]
\centering
\begin{center}
\includegraphics[scale=0.35]{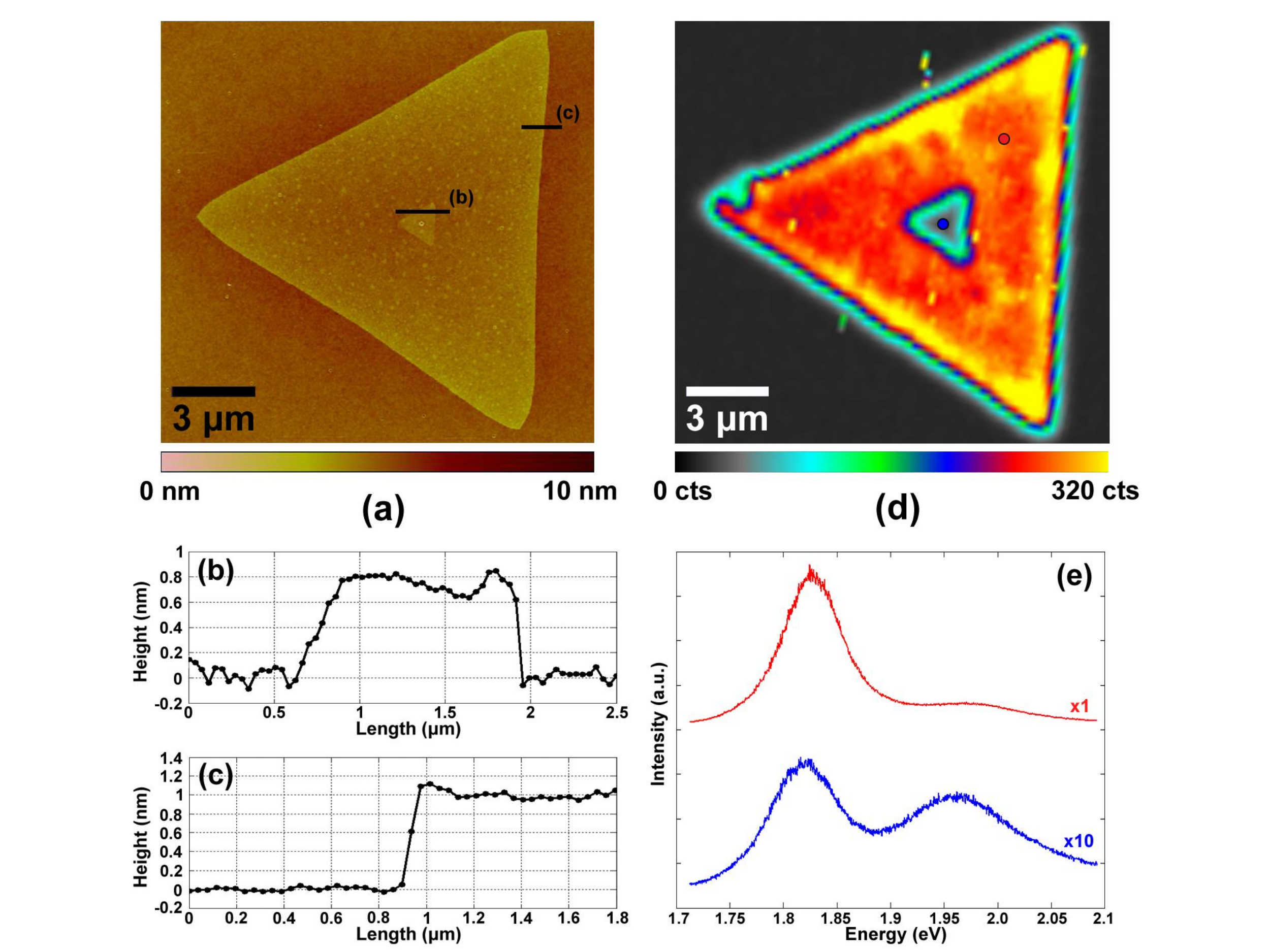}
\end{center}
\caption {AFM image of a typical MoS$_2$ crystal (a), step height measurement of the monolayer to bilayer step (0.7 nm) (b), step height measurement from substrate to monolayer (0.9 nm) (c), PL intensity map of the same MoS$_2$ crystal (d), and individual spectra for the monolayer (red) and bilayer (blue) regions (e).}
\end{figure}

\begin{figure}[h!]
\centering
\begin{center}
\includegraphics[scale=0.35]{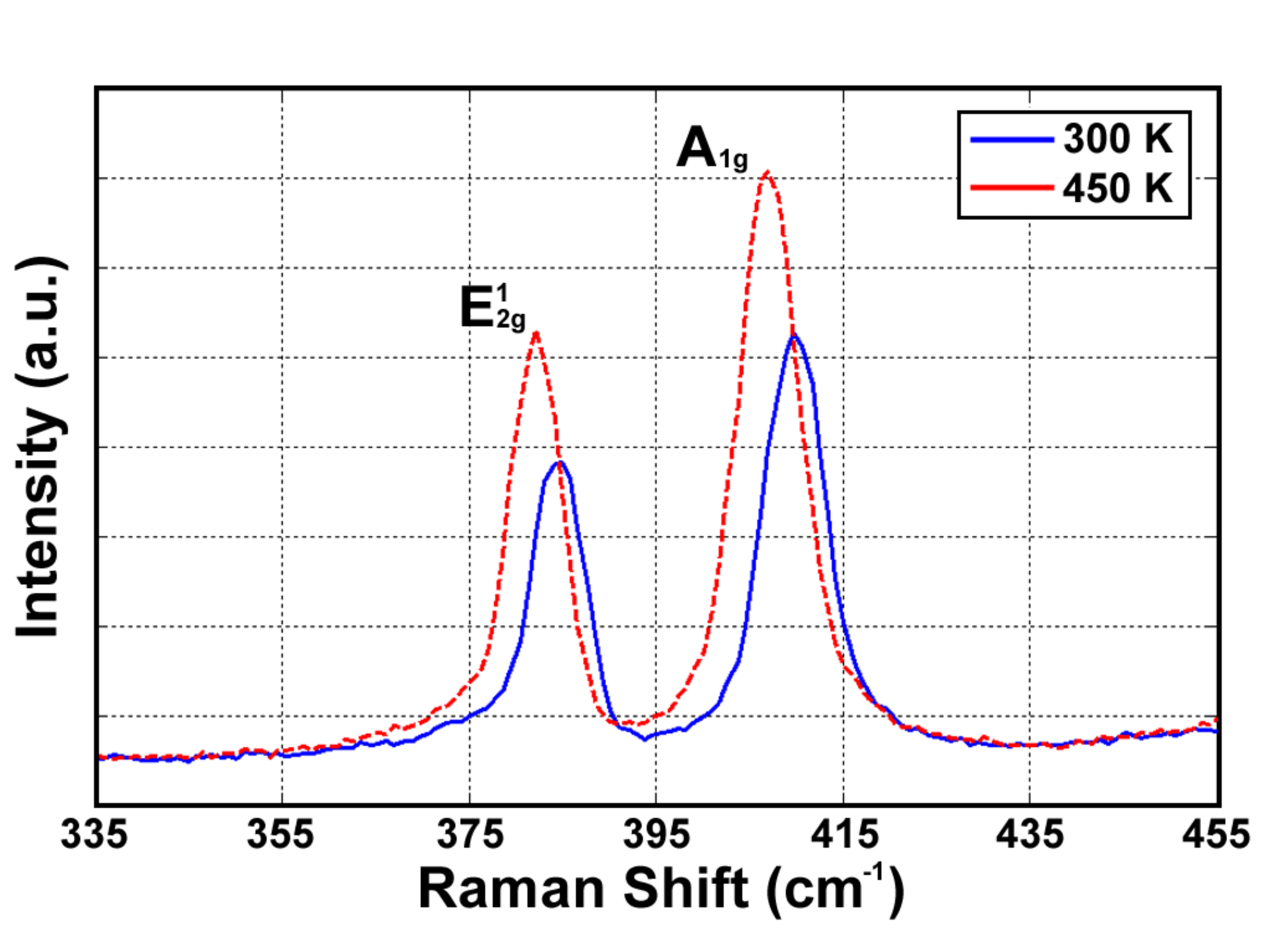}
\end{center}
\caption {Raman spectra of CVD monolayer MoS$_2$ taken at 30$^\circ$C and 175$^\circ$C plotted on the same scale showing the position of the E$^1$$_{2g}$ in-plane and A$_{1g}$ out-of-plane modes as well as an anomalous increase in intensity of the Raman signature.}
\end{figure}

\begin{figure}[h!]
\centering
\begin{center}
\includegraphics[scale=0.35]{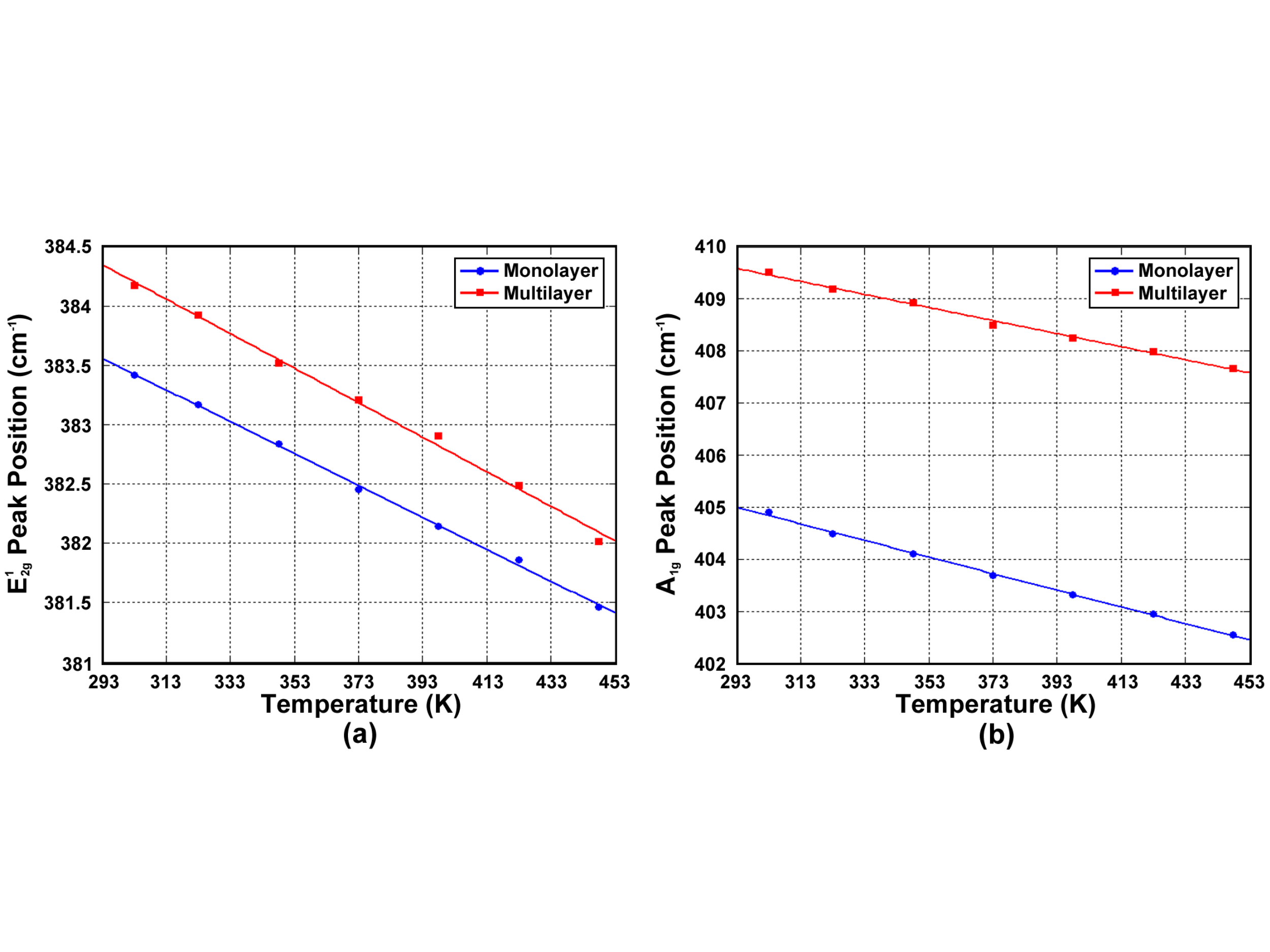}
\end{center}
\caption {Position of the E$_{2g}$ (a) and A$_{1g}$ (b) peaks as a function of temperature for CVD monolayer (red square) and exfoliated multilayer (blue circle) MoS$_2$.}
\end{figure}

\subsection{Theory} 
Any temperature-dependent changes in the calculated phonon density of states is due to the anharmonic terms in the lattice potential energy. In contrast to strictly harmonic or quasi-harmonic calculations, anharmonic frequency shifts are the result of coupling between phonons having different momentum $\textbf{q}$ and band index $j$, as well as thermal expansion of the lattice. The calculated phonon density of states for molecular dynamics simulations at representative temperatures of 50 K and 300 K are shown in \textbf{Figure 4}. 

\begin{figure}[h!]
\centering
\begin{center}
\includegraphics[scale=0.35]{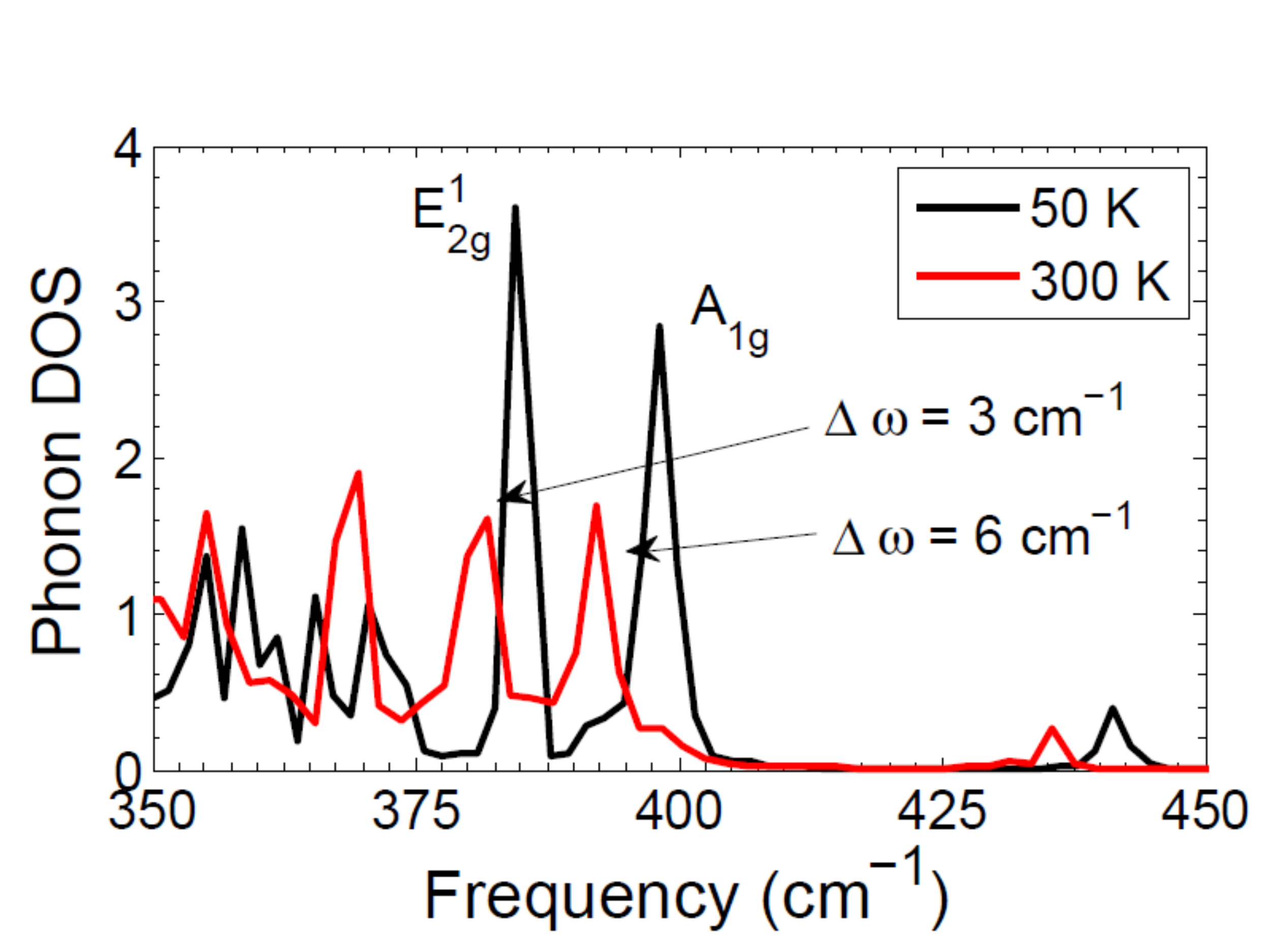}
\end{center}
\caption {The calculated phonon density of states for monolayer MoS$_2$ at temperatures of 50 K and 300 K.}
\end{figure}

At both temperatures, the Raman-active $A_{1g}$ and $E_{2g}$ peaks are well-defined, although slightly redshifted in frequency relative to estimates made in the harmonic approximation\cite{molina2011phonons}. It is clear that as the temperature is increased by several hundred Kelvin, both peaks shift toward lower frequencies, and that the frequency shift of the $A_{1g}$ peak is more pronounced than that of the $E_{2g}$ peak. 

This temperature dependent shift of both peaks is due to the anharmonic contributions to the interatomic potential energy, mediated by phonon-phonon interactions. Intuitively, it makes sense that the out-of-plane $A_{1g}$ mode shows a stronger temperature response in the monolayer, where there are no weak interlayer interactions restricting the vibrations away from the basal plane. While thermal expansion of the crystal is technically considered an anharmonic effect, it is a distinct physical phenomenon from the anharmonic coupling between phonons\cite{calizo2007temperature}. Because the size of the supercell is held fixed during simulations at different temperatures, we expect thermal expansion to play a minimal role since the in-plane structure does not change appreciably. 

We note that the $A_{1g}$ mode has a stronger coupling to electrons than the $E_{2g}$ mode, which could explain the larger frequency shift of the phonons. We can compare this to the case of superconducting MgB$_2$, in which the anharmonic renormalization is greatest for the modes that couple most strongly to electrons\cite{lazzeri2003anharmonic}.

\begin{table}[!h]
\begin{tabular}{ l l l l l}
\hline \hline 
 & & $X_{B}$ & $X_{ML}$ & $\Delta_{ML}$ \\  
 \hline 
$A_{1g}$   & & -0.013  & -0.016    & 6 cm$^{-1}$ \\ 
$E_{2g}$   & & -0.015  & -0.013     & 3 cm$^{-1}$ \\ 
\hline \hline 
\end{tabular}
\caption{Comparison of the experimentally-determined temperature coefficients (in units of cm$^{-1}$/K) for both bulk and monolayer MoS$_2$ as well as the frequency shifts calculated via molecular dynamics for the monolayer.} 
\end{table}

%%%%%%%%%%%%%%%%%%%%%%%%%%%%%%%%%%%%%%%%%%%%%%%%%%%%%%%%%%%%%%%%%%%%%%%%
\section{Conclusion}
In summary, we have used a combination of Raman spectroscopy and first-principles molecular dynamics simulations to study the temperature dependence of the Raman active peaks in monolayer MoS$_2$. We have investigated the temperature dependence of the E$_{2g}$ and A$_{1g}$ peaks in the Raman spectra of monolayer MoS$_2$ grown by CVD on Si/SiO$_2$ substrates. Micro-Raman spectroscopy was carried out using the 532 nm laser excitation over the temperature range from 30 to 175$^\circ$C. The extracted values of the temperature coefficient of these modes are X = -0.013 cm$^{-1}$/$^\circ$C and X = -0.017 cm$^{-1}$/$^\circ$C for monolayer, and X = -0.015 cm$^{-1}$/$^\circ$C and X = -0.013 cm$^{-1}$/$^\circ$C for bulk, respectively. Simulation results agree qualitatively with experiment, predicting a larger temperature-dependent shift for the A$_{1g}$ than for the E$_{1g}$ mode. 

We see that both the $A_{1g}$ and $E_{2g}$ peaks are redshifted with increasing temperature, and the magnitude of the $A_{1g}$ temperature coefficient is larger than that of the $E_{2g}$ mode. The temperature dependent shift of both peaks is attributed to anharmonic contributions in the interatomic potential energy due to phonon-phonon interactions. The larger anharmonic frequency shift of the $A_{1g}$ mode is due to the stronger electron-phonon coupling relative to other Raman-active modes. Future work will look at the temperature effects of few-layer MoS$_2$ as well as monolayers containing Mo defect sites. 

%Computational Center for Nanotechnology Innovations (CCNI).
%\end{acknowledgments}

%\bibliography{MoS2}
\bibliography{MoS2}
\end{document}